\documentclass{raa_twocolumn} 

\usepackage{graphicx,times}             
\usepackage{natbib}
\usepackage{amssymb,amsmath}
\usepackage{textcomp} 
\bibpunct{(}{)}{;}{a}{}{,}
\usepackage[pagebackref=true]{hyperref}

\usepackage{xcolor}
\usepackage{tcolorbox} 
\tcbuselibrary{breakable} 
\newif\ifshowcomments
\showcommentstrue 

\ifshowcomments
  \newcommand{\highlightcomment}[2][yellow!20]{%
    \begin{tcolorbox}[
      colback=#1,            
      colframe=#1,           
      boxrule=0pt,           
      arc=0pt,               
      left=2pt,right=2pt,    
      top=1pt,bottom=1pt,    
      breakable              
    ]
    #2
    \end{tcolorbox}%
  }
\else
  \newcommand{\highlightcomment}[2][yellow!20]{} 
\fi

\usepackage{xcolor} 

\begin{document}

  \title{COLIBRÍ (SVOM/FM-GFT): Instrumentation and Performances on the SVOM Alerts}

   \volnopage{Vol.0 (202x) No.0, 000--000}      
   \setcounter{page}{1}          

 \author{
S. Basa\inst{1,2}
\and W.~H.~Lee\inst{3}
\and A.~M.~Watson\inst{3}
\and F.~Dolon\inst{2}
\and J.~Floriot\inst{1}
\and J.-L.~Atteia\inst{4}
\and D.~Dornic\inst{5}
\and E.~E.~Lugo-Ibarra\inst{6}
\and L.~Figueroa\inst{6}
\and R.~Langarica\inst{3}
\and H.~Valentin\inst{4}
\and M.~Ageron\inst{5}
\and F.~Agneray\inst{1}
\and L.~C.~Álvarez~Núñez\inst{3}
\and C.~Angulo-Valdez\inst{3}
\and S.~Antier\inst{7}
\and T.~Auphan\inst{5}
\and M.~Baumann\inst{8}
\and L.~Bautista\inst{4}
\and R.L.~Becerra\inst{3}
\and S.~Benahmed\inst{4}
\and H.~Benamar\inst{5}
\and C.~Blanpain\inst{2}
\and O.~Boulade\inst{8}
\and Y.~Bounab\inst{8}
\and J.~Boy\inst{4}
\and N.R.~Butler\inst{9}
\and E.~O.~Cadena~Zepeda\inst{6}
\and S.~Cuevas\inst{3}
\and A.~de~Ugarte~Postigo\inst{1}
\and C.~Delisle\inst{8}
\and M.~Devigny\inst{4}
\and J.G.~Ducoin\inst{5}
\and F.~Fortin\inst{4}
\and J.~Fuentes-Fernández\inst{3}
\and C.~Gaïti\inst{4}
\and L.~García-García\inst{6}
\and P.~Gallais\inst{8}
\and R.~Gill\inst{10}
\and N.~Globus\inst{6}
\and G.~Guisa\inst{6}
\and E.~Kajfasz\inst{5}
\and D.~Lafforgue\inst{4}
\and A.~Langlois\inst{4}
\and M.~Larrieu\inst{4}
\and J.~Landa\inst{6}
\and J.~Lecubin\inst{2}
\and D.~López-Cámara\inst{11}
\and E.~López~Angeles\inst{6}
\and S.~Lombardo\inst{1}
\and F.~Magnani\inst{5}
\and N.~Mandarakas\inst{1}
\and A.~Malgoyre\inst{2}
\and R.~Mathon\inst{4}
\and E.~Moreno~Méndez\inst{12}
\and C.~Moreau\inst{1}
\and A.~Nouvel~de~la~Flèche\inst{4}
\and J.~L.~Ochoa\inst{6}
\and L.~Ortiz\inst{6}
\and M.~H.~Pedrayes-López\inst{6}
\and M.~Pereyra\inst{6,13,12}
\and L.~Provost\inst{8}
\and P.~Ramon\inst{4}
\and N.A.~Rakotondrainibe\inst{1}
\and S.~Ronayette\inst{8}
\and J.~Ruiz~Díaz-Soto\inst{3}
\and F.~Sánchez~Álvarez\inst{3}
\and B.~Schneider\inst{1}
\and A.~Secroun\inst{5}
\and N.~Striebieg\inst{4}
\and S.~Tinoco\inst{3}
\and M.~Tourner-Sylvain\inst{5}
\and F.~Valenzuela\inst{6}
\and D.~Vincent\inst{4}
}

\institute{
Aix Marseille University, CNRS, CNES, LAM, Marseille, France\\
    \and
Observatoire des Sciences de l’Univers (OSU) Institut Pythéas, Aix-Marseille Université – CNRS – IRD – INRAE, Aix-en-Provence / Marseille, France\\
    \and
Instituto de Astronomía, Universidad Nacional Autónoma de México, Apdo. Postal 70-264, CDMX, Mexico\\
    \and
IRAP, Université de Toulouse, CNRS, CNES, UPS, 31401 Toulouse, France\\
    \and
Aix Marseille Univ, CNRS/IN2P3, CPPM, Marseille, France\\
    \and
Instituto de Astronomía, Universidad Nacional Autónoma de México, Apdo. Postal 106, 22800 Ensenada, Baja California, Mexico\\
    \and
IJCLab, Univ Paris-Saclay, CNRS/IN2P3, Orsay, France\\
    \and
CEA Paris-Saclay, Irfu/Departement d’Astrophysique, 9111 Gif-sur-Yvette, France\\
    \and
School of Earth and Space Exploration, Arizona State University, Tempe, AZ 85287-1404, USA\\
    \and
Instituto de Radioastronomía y Astrofísica, UNAM, Morelia, Michoacán, México\\
    \and
Instituto de Ciencias Nucleares, UNAM, CDMX, Mexico\\
    \and
Facultad de Ciencias, Universidad Nacional Autónoma de México, CDMX, Mexico\\
    \and
Secretaría de Ciencia, Humanidades, Tecnología e Innovación (SECIHTI), Mexico\\
        \vs\no
   {\small Received 202x month day; accepted 202x month day}}

\abstract{ 
COLIBRÍ, the French–Mexican Ground Follow-up Telescope (FM-GFT) for SVOM, is a 1.3-meter rapid-response optical facility specifically developped for prompt, multi-band observations of GRB afterglows and for delivering sub-arcsecond localizations of optical counterparts for detailed follow-up studies. The telescope operates through a fully automated system that manages the entire workflow—from alert reception to counterpart identification. Commissioning results confirm that the telescope meets design specifications, and this paper presents a comprehensive performance assessment of the system’s capabilities.
\keywords{transients:Gamma-ray bursts --- telescopes --- instrumentation: photometers --- methods: observational – techniques: image processing}
}

   \authorrunning{S. Basa, W.H. Lee, A.M. Watson, F. Dolon at al.}            
   \titlerunning{COLIBRÍ (SVOM/FM-GFT): Instrumentation and Performances}  

   \maketitle
%
%

\section{Introduction}           

\subsection{COLIBRÍ in the Context of SVOM}
Cosmic explosions are among the most spectacular and informative events in the Universe. They not only illuminate the most distant and extreme regions of the cosmos but also provide unique laboratories for studying fundamental physics. Today, these phenomena serve as standard candles for measuring cosmic expansion (Type Ia supernovae), as probes of the early Universe (Gamma-Ray Bursts, GRBs), as sites of extreme relativistic physics, as direct witnesses to the birth of compact objects such as black holes and neutron stars and as probes of cosmic nucleosynthesis. The rise of multi-messenger astronomy, combining electromagnetic, neutrino, and gravitational-wave observations, has further amplified their scientific value.

GRBs are especially remarkable: they are the most energetic explosions since the Big Bang,  typically releasing with isotropic-equivalent-energy release in the range $10^{48}$ erg to $10^{55}$ erg within minutes. They originate either from the deaths of massive stars or from the mergers of compact stellar remnants. Their extreme luminosity allows them to probe regions of the Universe and physical regimes otherwise inaccessible to observation.

The Sino-French SVOM (Space-based multi-band astronomical Variable Objects Monitor) mission is designed to capitalize on this potential \citep{Wei+etal+2016}. By combining space-based instruments and ground-based telescopes, SVOM detects GRBs, localizes them from arcminutes to arcseconds, studies their prompt emission, and ensures rapid optical follow-up \citep{Cordier+etal+2026a}. Its ground segment distributes alerts within minutes, refines localization to sub-arcsecond precision, and identifies high-redshift candidates ($z>6$) \citep{Louvin+etal+2026} \citep{Huang+etal+2026}. Launched in June 2024 from Xichang, China, SVOM is planned for a three-year primary mission with at least a two-year extension.

Two robotic Ground Follow-up Telescopes (GFTs), managed by France, Mexico \citep{basa:hal-03838591} and China \citep{Wu+etal+2026}, are an integral part of the mission. They:
\begin{itemize}
\item Observe GRBs in visible and infrared, in the specific case of the COLIBRÍ/FM-GFT, from minutes to at least one day after the alert.
\item Provide sub-arcsecond localization within 5 minutes, complementing the $\sim$26 arcminute onboard precision of SVOM/ECLAIRs.
\item Identify low signal-to-noise triggers that do not prompt satellite slews, particularly very high-redshift GRBs.
\item Detect “dark GRBs” observable in gamma or X-rays but not in optical wavelengths.
\item Serve as a bridge to major facilities such as GTC (Gran Telescopio Canarias), NOEMA (NOrthern Extended Millimeter Array), NOT (Nordic Optical Telescope), NTT (New Technology Telescope), and VLT (Very Large Telescope) by providing rapid preliminary photometric redshift estimates.
\end{itemize}

In order to fulfill its scientific requirements, COLIBRÍ, the official name of the GFT managed by France and Mexico, has very specific requirements (Tab.~\ref{tab:specs}): high availability for alert observations, very good sensitivity, fast pointing speed (on target in under 20 seconds after the alert reception), multiband photometric capabilities (from 400 to 1800 nm, with three channels capable of simultaneous imaging), and a field of view covering the SVOM/ECLAIRs trigger error box (26\arcmin).

\begin{table*}
\begin{center}
\caption[]{Main requirements on the COLIBRÍ observatory.}
\label{tab:specs}
\begin{tabular}{ll}
\hline\noalign{\smallskip}
Observatory location &  Observatorio Astronómico Nacional, Sierra de San Pedro Mártir, Mexico \\
Delay for pointing &   $<$ 20 seconds \\
Precision of localization &  $<$ 0.5 arcsec \\
Primary mirror diameter & 1.3 m  \\
Photometric channels & Three simultaneous channels: two in the visible ($Bgri$ and $zy$) and one in the NIR ($JH$)  \\
Field of view & 26 arcmin in the visible and 22 arcmin in the NIR\\
Pixel scale & 0.38 arcsec/pix in the visible and 0.64 arcsec/pix in the NIR  \\
Real-time data processing & $<$ 5 minutes  \\ 
\noalign{\smallskip}\hline
\end{tabular}
\end{center}
\end{table*}

\subsection{COLIBRÍ, an Astronomical Observatory for the Transient Sky}
COLIBRÍ is the outcome of a close collaboration between France and Mexico, with additional participation from Arizona State University. Institutional partners include Aix--Marseille University (AMU), CNES, and CNRS in France, and UNAM and SECIHTI in Mexico. The project is supported by several laboratories in France (in alphabetical order: CEA, CPPM, IRAP, LAM, OHP/OSU~Pytheas, and OSU~Pytheas) and in Mexico (the Instituto de Astronomía in Mexico City, CDMX, and Ensenada, BC). This partnership was formalized through a Memorandum of Understanding signed in~2018.

The project is designed to support the SVOM mission, but it also aims to go beyond this by studying the transient sky in a broader context. The ongoing decade is indeed marked by the arrival of several large-scale facilities dedicated to time-domain astronomy (VRO, CTA, SKA, LOFAR, etc.), complemented by all-sky detectors for neutrinos (KM3NeT, IceCube) and gravitational waves (Advanced Virgo, LIGO and KAGRA). These instruments provide alerts within seconds to minutes, creating unprecedented opportunities to investigate the very first stages of cosmic explosions. 

By responding quickly to all these alerts, with an excellent sensitivity and a large wavelength coverage, COLIBRÍ contributes uniquely to the study of the Universe’s most energetic transient phenomena.

\subsection{The COLIBRÍ Observatory}
The telescope is installed at the Observatorio Astronómico Nacional (OAN) in the Sierra de San Pedro Mártir, Baja California \citep{2007RMxAC..31...47T}, at the observatory’s highest point ($31^\circ02\arcmin41\farcs52$ N $115^\circ27\arcmin52\farcs13$ W and an altitude of $\sim$2940 m). 

OAN’s suitability has been confirmed through decades of operation and independent evaluations \citep[e.g.,][]{Sch_ck_2009}. These studies show that the median seeing is 0\farcs8 and that ~80\% of nights are observable and 60\% photometric, demonstrating that the site is similar to other world-class locations such as Mauna Kea, La Silla, Cerro Tololo, and La Palma.

The facility consists of:
\begin{itemize}
\item an enclosure building containing the observing room and control room;
\item a service building for power and cooling infrastructure,
\item a 1.3-m alt-az telescope with a wide-field imager operating in the visible and the near-infrared;
\item a software robotic control system managing operations;
\item a control center managing data processing and archiving.
\end{itemize}

The details of all these components, as well as the performance on SVOM alerts, are presented in the following sections of this paper.

\section{Details of the system constituting the COLIBRÍ Observatory}

\subsection{Enclosure and Infrastructure}

\begin{figure*}
    \centering
    \includegraphics[width=0.9\textwidth]{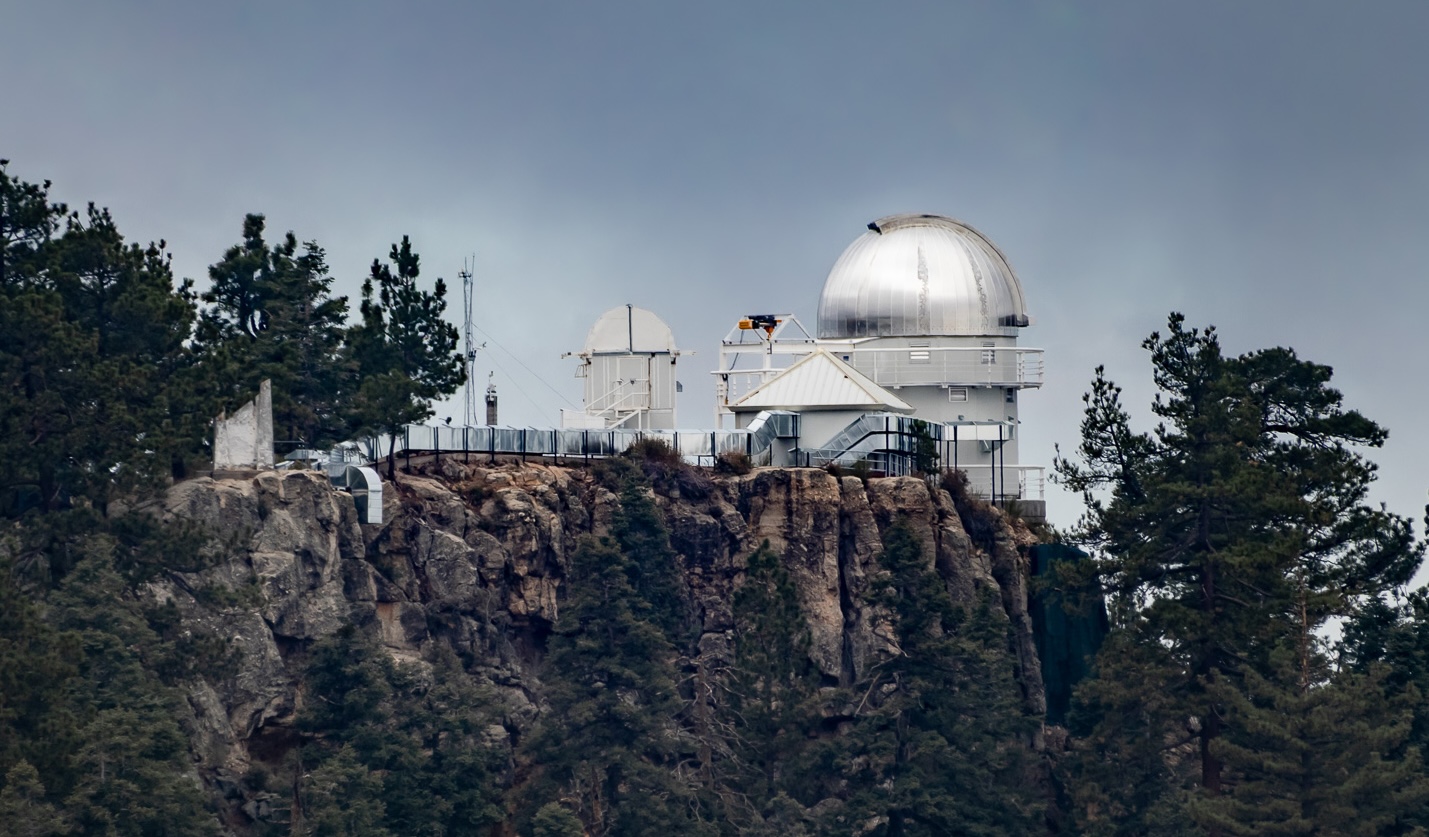}
    \caption{The COLIBRÍ buildings. On the right is the main building with an aluminum dome. The building just to the left of this, with the sloped roof, is the service building. The exhausts for the air-conditioning system extend along the edge of the cliff to the left. The building in the middle, with the clam-shell enclosure, is for the BOOTES-5 telescope. The weather station and seeing monitor are on the column and mast to the left of the BOOTES-5 building.}
    \label{figure:infrastructure}
\end{figure*}

COLIBRÍ is housed in two purpose-built buildings at the highest point of the OAN ((see Fig.~\ref{figure:infrastructure})). 

The buildings have steel frames, concrete floors, and are sheathed with composite panels. The upper floor of the main building contains the telescope enclosure and observing room, and the lower floor contains the control room. The telescope is supported by a reinforced concrete column that is embedded 2.10 meters into the underlying rock and passes through the control room into the observing room. The adjacent service building accommodates transformers and air-conditioning units. Both buildings are connected to the same physical electrical-grounding system. 

The detailed design of the buildings and associated infrastructure is described in \cite{spie-2024-lugo}, but we summarize some of the most important points here. The building design and construction were supervised by the Ensenada branch of the Instituto de Astronomía.

\subsubsection{Thermal Control}

We paid a very careful attention to thermal issues in order to preserve the excellent image quality of the site. The dome has a bare aluminum surface, which has low emissivity. The dome is lined with 15~cm of polystyrene foam and the walls of the buildings are formed from 15~cm insulating panels, both of which provide excellent thermal insulation. The walls of the observing floor have twelve louvers (with a total area of 5 m$^2$), which facilitate nocturnal ventilation. The heat generated within the dome is limited to the dome controller, the safety PLC, and the instrument electronics, and all have been carefully optimized. 

The control room contains all of the control computers, the telescope control cabinet, the communications equipment, and the instrument power supplies and cryogen compressors. The control room is air conditioned and is thermally isolated from the observing room by a false roof of 7.5~cm insulating panels. The space above the false roof and below floor of the observing room is actively ventilated to mitigate any remaining heat leaks. The underside of the concrete-and-steel floor of the observing room is insulated with 2.5~cm insulating panels for the same reason.

The service building contains the air-conditioning units for the control room. The warm air generated by these units and by the transformers in the service building is vented approximately 60 meters to the south east, about half-way to the 2.1 meter telescope. This direction is perpendicular to the two prevailing wind directions, so the warm air does not blow back onto the telescope (or onto other nearby telescopes, which are aligned along a line to the south west). The air-conditioning system does not directly cool the observing room, but it allows us to thermally isolate the control room from the dome and exhaust the heat generated in the control room far from the telescope, permitting better passive control of the environment in the observing room. Actively cooling the observing room to the night-time temperature would have required a considerably more powerful air-conditioning system, with concomitant higher costs.

Our thermal design has proven to be successful. We typically find that at the start of the night, the delivered images seen in the two channels of DDRAGO are between 0.8 and 1.0 arcsec, with the lower limit set by the pixel size of 0.38 arcsec.

\subsubsection{Dome}

The telescope’s optically fast design makes it possible to house it in a compact 7.5\,m dome, manufactured jointly by the German company ASTELCO and the Italian company Gambato. To avoid the dome inducing vibrations in the telescope, the main building has a separate foundation to the central column, the floors are isolated from the column by neoprene, and the vertical steel columns of the main building are filled with concrete.

To ensure that the telescope’s rapid response is not limited by the dome speed, we commissioned a custom rotation mechanism. The dome has two speeds. At the faster speed, it can rotate through 180 degrees in 13 seconds. At the normal speed, it takes about twice as long. The torques associated with the faster speed required careful attention to the mechanical structure of the main building. We expect to commission the faster speed during the winter of 2025--2026.

\subsubsection{Electrical Supply}

The electrical supply for the telescope and buildings is complicated by the requirement that the telescope be supplied with three-phase 230/400 V 50 Hz, whereas the standard power supply at the observatory is 120/220 V 60 Hz. We supply the computers, dome, and other equipment with standard 60 Hz power via a standard UPS, and in parallel we supply a separate UPS with 400 V 60 Hz, which converts it to 230/400 V 50 Hz for the telescope.

\subsubsection{Instrument and Mirror Handling}

One disadvantage of the compact dome is that the space available for handling the mirrors and instruments is very confined; the distance between the concrete column and the inner edge of the building steel structure is only about 2.1 meters. This required a number of innovative solutions, which are described briefly here and in detail in \cite{spie-2022-pedrayes} and \cite{spie-2024-pedrayes}. All of these tools were designed and manufactured in Ensenada branch of the Instituto de Astronomía.

Perhaps the most difficult issue is transporting the primary mirror to a site elsewhere on the mountain for aluminization. This will be achieved by pointing the telescope to the zenith, installing rails under the primary, raising a cart under the cell to take its weight and allow it to be decoupled from the mount, lowering the cart, and then moving it on the rails onto an elevating platform, with a surface of approximately 1.8 by 1.8 meters, between the telescope and the doors to the outer level. Next, it is lowered to the level of the observing floor and then rolled on rails out onto the external platform. After this, the primary mirror is lowered onto a vehicle with the external gantry crane.

\begin{figure*}
    \centering
    \includegraphics[width=0.9\textwidth]{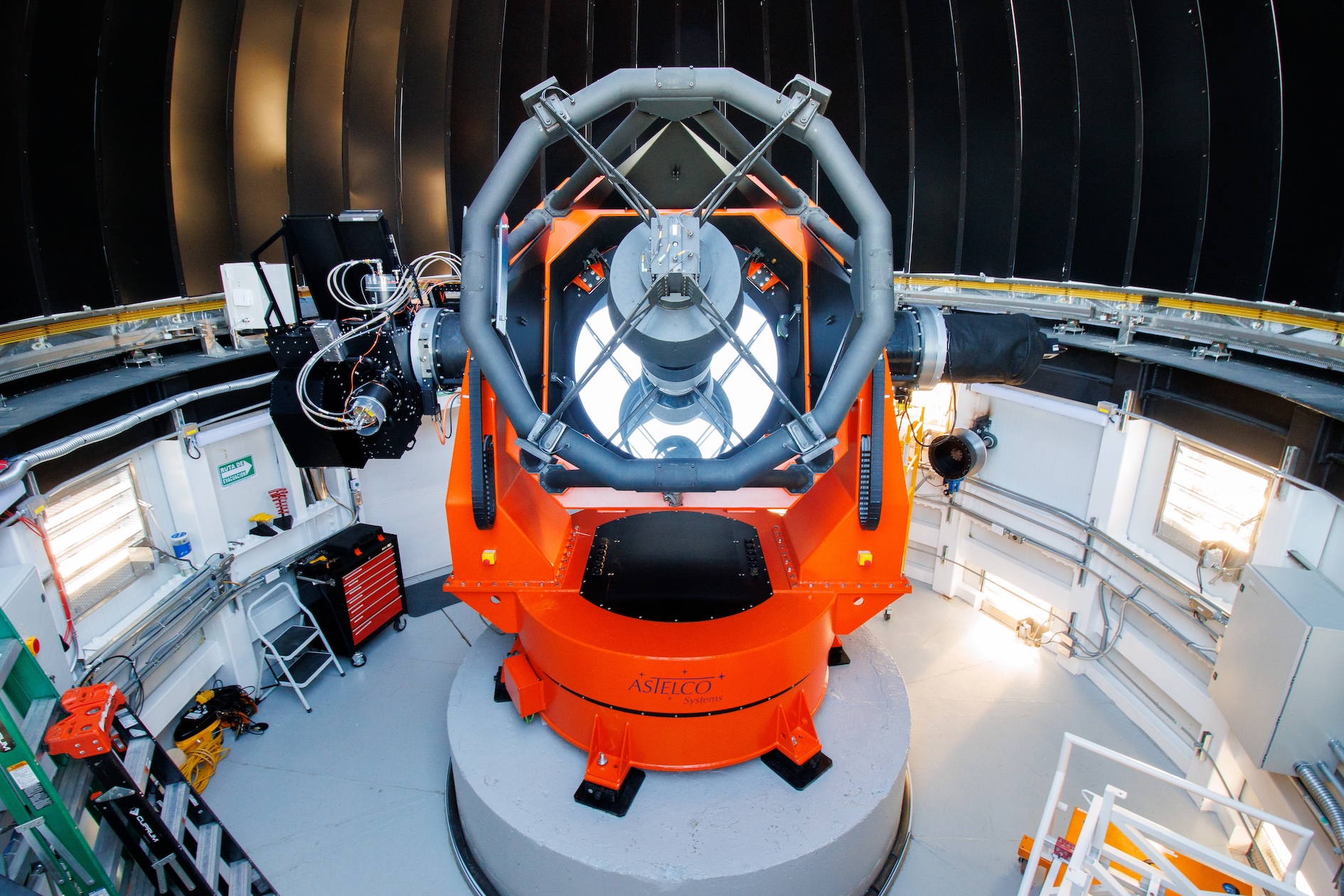}
    \caption{The COLIBRÍ telescope in its enclosure with the DDRAGO camera (left) and the OGSE test camera (right) mounted on its Nasmyth ports. DDRAGO is rotated to show the blue (upper) and red (lower)  CCDs. The dummy derotator, used for balancing the instruments, is seen on the wall just to the right of the telescope.}
    \label{figure:colibri}
\end{figure*}

The design of the telescope derotators requires us to not exceed a torque of 100 N~m and does not permit us to balance instruments on the real derotators. Therefore we have installed a dummy derotator, essentially the same flange on an axle with a brake, on the outer steel structure in the observing room (see Fig.~\ref{figure:colibri}).

The last difficult issue was inserting the instrument into the derotator, as the mechanical tolerances between the instrument and derotator flanges are very tight. The weight of the instrument is 330 kg, which does not make this task any easier. Therefore, we designed a cart with five degrees of freedom with fine control and a screw-driven insertion mechanism. The cart is also used to support the instrument during maintenance, as it allows us to remove parts without exceeding the torque limits of the derotator.

\subsection{Telescope}
The telescope has an alt-azimuth mount with two Nasmyth foci equipped with two derotators. The telescope has been built by the German company ASTELCO and the two main mirrors were polished by the French companies AstroOptique Cardoen for the M1 and Winlight-Bertin for the M2. The main characteristics are summarized in Tab.~\ref{tab:telescope}.

The project made the choice to have an unprotected coating (pure aluminium) for the two main mirrors, M1 and M2, of the telescope. This choice was driven by the fact that the observatory has equipment on site to carry out periodic recoatings (every two years). A complete set of tools has then been developed to carry out this task every two years (\citep{spie-2022-pedrayes,spie-2024-pedrayes}.

\begin{table}
\begin{center}
\caption[]{Main characteristics of the telescope.}
\label{tab:telescope}
 \begin{tabular}{ll}
\hline\noalign{\smallskip}
Delay for pointing &   $<$ 20 seconds  \\
Maximum settle time &   $<$ 1 second  \\
Minimum elevation  &   15°  \\
Absolute localization accuracy &   $<$ 2.5 arcsec RMS   \\
Maximum load on the derotator &   350 kg  \\  
\noalign{\smallskip}\hline
\end{tabular}
\end{center}
\end{table}

\subsection{DDRAGO}

\begin{figure*}
    \centering
    \includegraphics[height=0.52\textwidth]{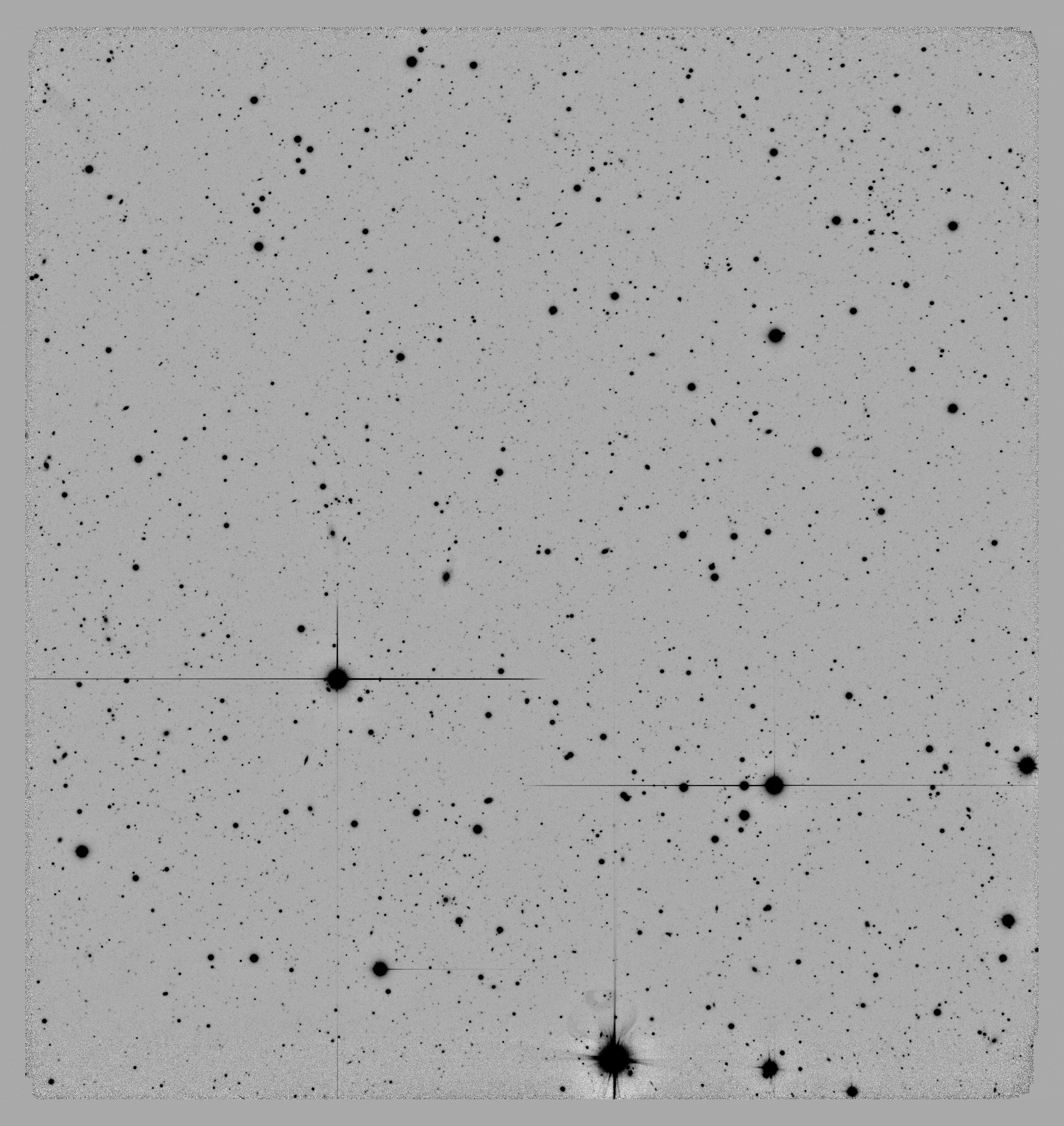}
    \includegraphics[height=0.52\textwidth]{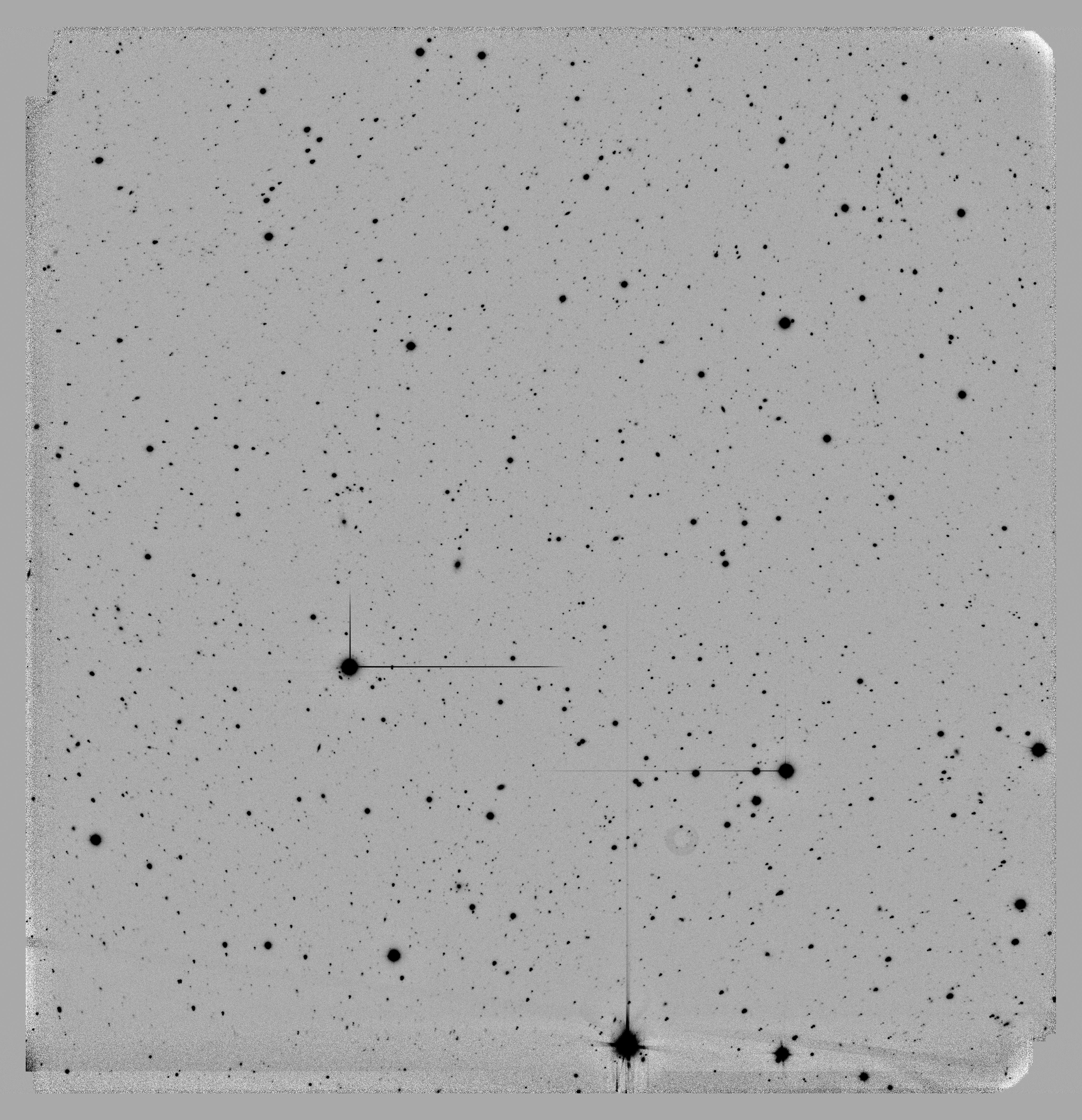}
    \caption{Simultaneous images with DDRAGO in the blue channel in $r$ (left) and in the red channel in $z$ (right) of the field of the SVOM GRB 251013C \citep{gcn-42222} taken on the night of 2025 October 14 and reported by \cite{gcn-42242}. The images are coadditions of ninety exposures each of 60 seconds, with dithering. The field is about 26{\arcmin} to a side. The $3\sigma$ limiting magnitudes are $r \approx 24.5$ and $z \approx 23.0$.}
    \label{figure:ddrago-images}
\end{figure*}

DDRAGO is one of the principal instruments for COLIBRÍ. It can be described as ``one and a half'' instruments: a two-channel, wide-field optical imager, complemented by the ambient-temperature reimaging optics that feed the CAGIRE infrared camera described below. 

Typical images with DDRAGO are shown in Fig.\ \ref{figure:ddrago-images}. DDRAGO is described in more detail in \cite{fuentes-2020}, \cite{spie-langarica-2022}, \cite{spie-farah-2022}, \cite{spie-angeles-2022}, \cite{spie-langarica-2024}, and \cite{spie-farah-2024}. DDRAGO was designed and largely manufactured in the Mexico City branch of the Instituto de Astronomía. Its name is taken from the phrase \emph{Detectando Destellos de RAyos Gamma en el Óptico} (Detecting Gamma-Ray Bursts in the Optical), which was chosen as a natural analog of that of its sister instrument CAGIRE. 

\subsubsection{Optics}

\begin{figure*}
    \centering
    \includegraphics[height=0.26\textwidth]{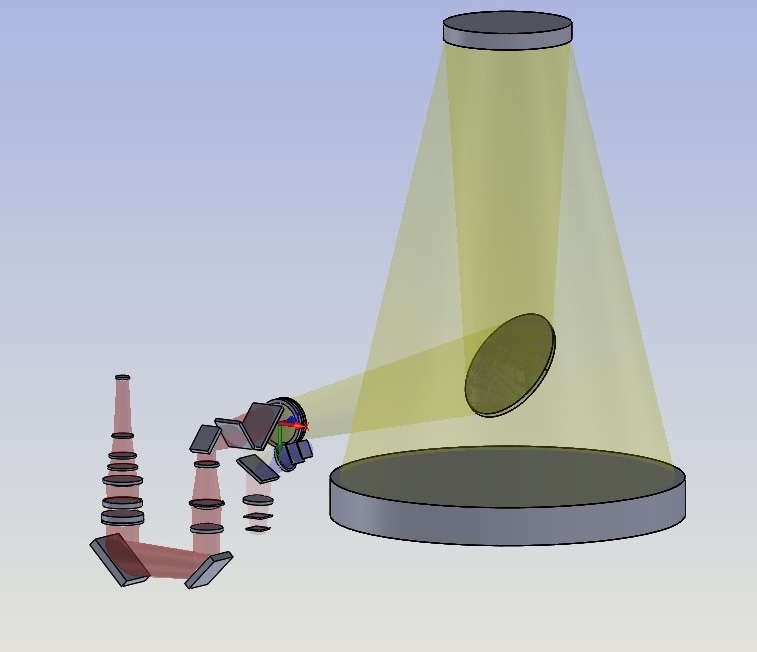}
    \includegraphics[height=0.26\textwidth]{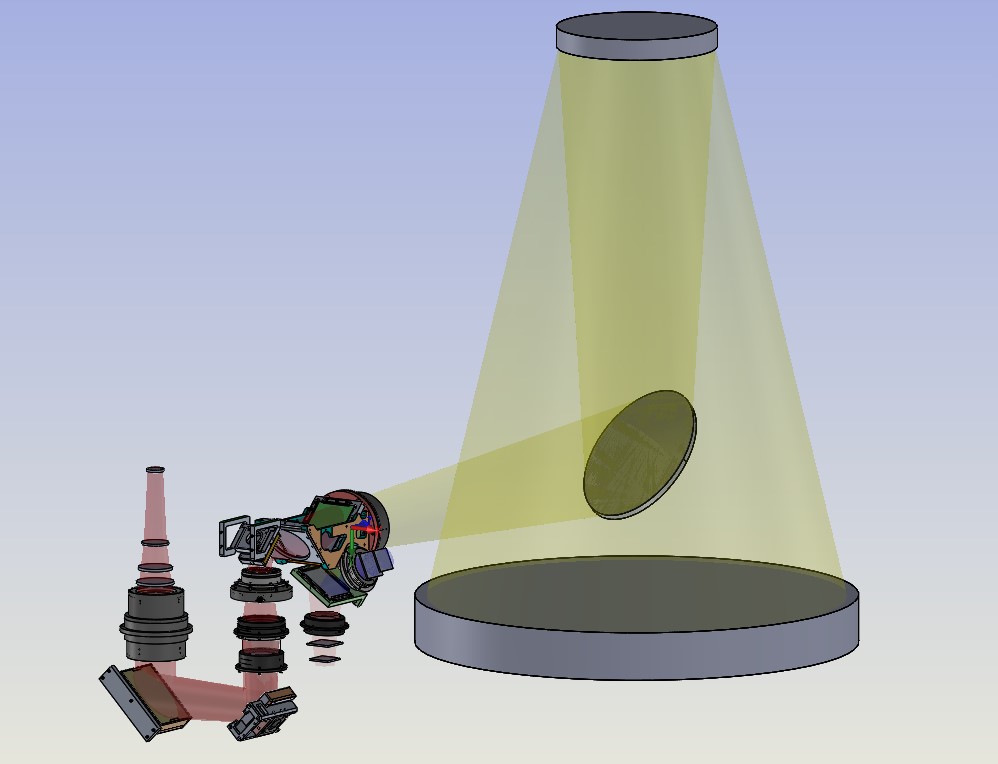}
    \includegraphics[height=0.26\textwidth]{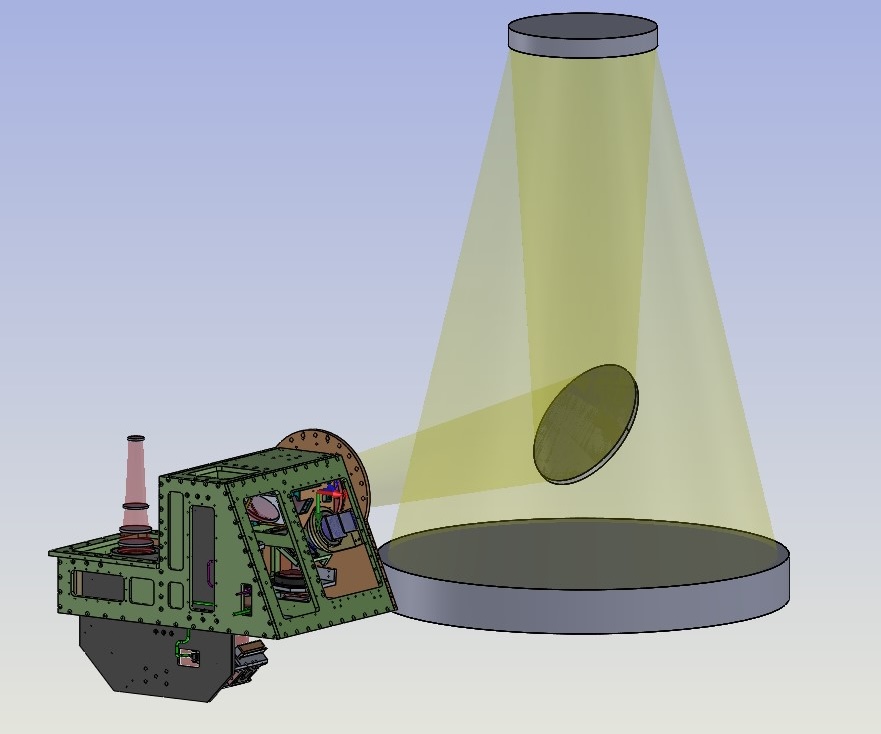}
    \caption{The DDRAGO optics and mechanics. The left panel shows only the optics. After being reflected from the tertiary mirror M3, the beam passes through the doublet L1L2. It then encounters the first dichroic D1 which transmits $JH$ light to the WOB and CAGIRE and reflects $grizy$ light to the CCDs. The optical light is further divided by the second dichroic D2 which transmits $zy$ light to the red CCD and reflects $gri$ light to the blue CCD. The middle panel shows the optomechanics. The right panel shows the support structure, with the access covers removed.}
    \label{figure:ddrago}
\end{figure*}

The optics of the telescope, the optical channels, and the infrared channels were designed and optimized simultaneously. The optics image the $g$, $r$, $i$ bands onto the blue detector, the $z$ and $y$ bands onto the red detector, and the $J$ and $H$ bands onto the CAGIRE infrared detector. The field of view is 25{\farcm}9 in the two optical channels and will be 22{\arcmin} in the infrared channel. The reimaging optics for CAGIRE are known as the Warm Optical Bench or WOB.

The optical design includes eleven lenses, two dichroics, one corrector plate, and three fold mirrors. The general distribution of the optics is shown in Fig.\ \ref{figure:ddrago}. After the common focal reducer (L1L2), the light encounters the first dichroic (D1) and the $grizy$ bands are reflected towards the CCDs, while the $JH$ bands are transmitted to the WOB. A second dichroic (D2), reflects $gri$ towards the blue CCD, and transmits $zy$ towards the red CCD. Each optical channel has a field lens (L3 and L4) in front of the detector, and filter wheels are located between L3 or L4 and the CCDs. These field lenses correct the field curvature from the telescope, and L4, tilted by 3 degrees, helps to correct the astigmatism introduced by D2. The final focal ratio at the CCDs is $f/6.3$. The two optical channels have tilt and piston adjustment to align the detectors with the two focal planes. After this adjustment, both channels are sufficiently confocal that they can be focused by adjusting the position of the telescope secondary mirror and do not need independent adjustment. The lenses, dichroics, and corrector plates for the optical channels were supplied by the U.S.\ company Custom Scientific and were verified in our laboratories.

The WOB supports a lens-based reimaging system which reduces the focal ratio from $f/6.3$ to $f/3.7$ for the infrared instrument CAGIRE. After the first dichroic (D1), a wedged corrector plate (CP) tilted in the opposite sense to D1 compensates the aberrations of D1 in transmission. The WOB has a total of 7 lenses, with one aspherical surface in one of the lenses, which act differently from a traditional collimator-camera system with an intermediate pupil plane. Instead, the design places the pupil plane cold stop after all the reimaging optics, avoiding all powered optical elements inside the cryostat except for a field lens with relaxed tolerances, thereby significantly reducing costs and risks. This concept was pioneered by the instrument RATIR \citep{butler-2012}, but has been implemented here with a faster beam and a larger field of view.

The original set of lenses for the WOB were supplied by the french company Trioptics. Unfortunately, when we inspected them after delivery, the coatings were inadequate. We requested that Trioptics remedy these defects, but did not obtain satisfaction. We have ordered a second set of lenses from the U.S.\ company Optimax, and anticipate the last lense will be delivered early 2026. DDRAGO is currently installed with a mass model in place of the these lenses and their optomechanics; we plan to manufacture, integrate, test, and install the real lenses before the end of 2026.

\subsubsection{Mechanics}

The main structure, shown in Figure \ref{figure:ddrago}, attaches the instrument to the Nasmyth station of the telescope. It also houses the optomechanical system and supports CAGIRE and the close electronics. It holds the detectors and their respective filters wheels and provides alignment mechanisms, and supports cables and hoses, the cable-wrap, and eyebolts for attaching the lifting harness. The structure has removable plates to allow access to the interior of the instrument for maintenance. It includes interface plates to the DDRAGO CCD detectors with tilt and piston adjustment. The structure uses plates of 7000-series aluminum alloy. We verified the performance of the structure by performing a flexure analysis at rotation angles from $-65$ to $+65$ degrees, corresponding to the full derotator range.

The optomechanical parts, shown in Figure \ref{figure:ddrago}, were manufactured mainly from 7000-series aluminum alloy. After manufacturing, metrology was performed with a Mitutoyo coordinate measuring machine. All the optomechanical parts have reference surfaces for the precise positioning of the optical elements and provide means for the fine centering of the lenses inside their barrels. Every barrel has a special machining to reduce scattered light: the inner diameter at the entrance acts as a baffle and the inner surface is grooved with a pattern of ridges. After centering, the lenses were preloaded to withstand transportation and handling shocks. Each optical element includes passive thermal compensators that reduce the effects of temperature changes on the optics over a wide range. 

We have divided the optomechanics between two reference surfaces. The D1 dichroic and CP corrector plate are installed in a substructure, and then this along with L1L2, L3, L4, and D2 are attached to the inner surface of the plate that forms the interface to the derotator. All of the optomechanics for the WOB will be installed in a single, removable plate.

All mechanical parts were black-finished internally with a mixture of type II anodyzing, Krylon® 1602 Ultra-Flat black paint, and Edmund Optics black flock paper.

\subsubsection{Filters}

Each channel has a the U.S.\ company FLI filter wheel for five 76 mm square filters. The blue channel has filters that are very close to Pan-STARRS $g$, $r$, and $i$,  a wide filter $gri$ that passes from the blue edge of $g$ to the red edge of $i$ (and is similar to Pan-STARRS $w$), and a $B$ filter. The red channel has filters that are very close to Pan-STARRS $z$ and $y$ and a wide filter $zy$ that passes from the blue edge of $z$ and as such is similar to SDSS $z$, although the response of our deep-depleted CCDs is quite different to those of the SDSS survey camera.

\subsubsection{Detectors}

Each optical channel has Spectral Instruments 1110S CCD detector systems, each with a backside-illuminated e2v 231-84 CCD. These detectors have a format of 4k × 4k with 15 micron pixels. The pixel scale is 0\farcs38 per pixel and the field size is 25\farcm9. The detectors have the “astro multi-2” AR coating and deep-depleted silicon. The red detector also has e2v’s fringing-suppression treatment. They are cooled by a Brookes Polycold® PCC Compact Cooler with PT-30 refrigerant. Each detector has an integrated Vincent/Uniblitz CS90HS1T0 shutter.

\subsubsection{Control}

The control system of DDRAGO has hardware in two locations: the control room and the close electronics cabinet mounted in DDRAGO structure. The hardware in the control room includes the two CCD service cabinets, which supply power and compressed PT-30 refrigerant to the CCDs, two Windows PCs to control the CCDs, two Linux PCs to run the rest of the robotic control system TCS, described below, network switches, and remote power distribution units. The hardware in the close electronics for DDRAGO include a USB extender, power supplies, and 1-wire environmental sensors. The USB extender is used for the filter wheels and the 1-wire adapter.

Cables, fibers, and hoses run between the two locations. The first are routed up to the ceiling of the control room. Then they pass along tunnels in the central concrete column to a central space under the telescope; the tunnels are sealed with polyurethane foam to avoid heat leaks. They then pass through the azimuth cable wrap up to the azimuth base and fork arms, and then through the elevation cable wrap up to the instrument.

\subsubsection{Image Quality}

The high-level requirement on the DDRAGO image quality is that the diameter that encircles 80\% of the energy ($d_{80}$) must be less than 0{\farcs}75 in all the $grizy$ filter. We measured the on-axis image quality using the $gri$ and the $zy$ filters prior to installation and found $d_{80}$ of 0{\farcs}58 in both channels. Estimates of the telescope image quality on its own give a $d_{80}$ of 0{\farcs}65, so currently the overall $d_{80}$ of the telescope plus instrument at the center of the field is 0{\farcs}87  and does not fulfill the requirement. Away from the field center, the image quality is further degraded by telescope aberrations.

We believe the telescope aberrations are astigmatism in M1 and comma from M1-M2 alignment and expect to correct them in Spring 2026. After this, we expect the $d_{80}$ of the telescope to be substantially smaller than the instrument and so expect to achieve an overall $d_{80}$ of about 0{\farcs}60 and thereby fulfill the requirement.

\subsection{CAGIRE}

CAtching Gamma-ray bursts InfraRed Emission (CAGIRE) is a near-infrared imager that will operate in the $J$ (1.2\,$\mu$m) and $H$ (1.6\,$\mu$m) bands \citep{2023ExA....56..645N}. 

Mounted on the Nasmyth focus of COLIBRÍ, on top of the DDRAGO camera, it is designed to image a field of view of 22$\times$22\arcmin, with a pixel size on the sky of 0.64\arcsec; this ensures that the camera will be limited by the sky background rather than internal radiation or readout noise. CAGIRE's field of view will thus entirely cover high-energy error boxes of GRB localization given by SVOM/ECLAIRs. Its unique combination of quick reactivity and near-infrared sensitivity will allow CAGIRE to address the issue of optically dark GRBs, which could be associated to events located at very high-redshift (z$>$6) and/or within massive host galaxies that absorb optical radiation through interstellar reddening.

The sensor, a 2k$\times$2k large-format astronomical array (ALFA), was manufactured by the French company Lynred. The ALFA sensor is loaned by ESA to the LabEx FOCUS, which provides it to CAGIRE and is, together with CNES, the main source of funding for the instrument.

\subsubsection{ALFA sensor, the heart of the camera}

The ALFA sensor was developed and produced by the French company Lynred. It is a large array format of 2k$\times$2k, 15\,$\mu$m wide photosites sensitive within the 0.8 -- 2.1\,$\mu$m range. Operating at 100\,K, its quantum efficiency in that range is greater than 70\%. The HgCdTe (MCT) photovoltaic sensor is grown on a CdZnTe substrate. The SFD readout circuit (Source Follower per sensor), alongside the coupling of each pixel to a MOS transistor, allows the non-destructive reading during acquisition. In this Up-The-Ramp mode (UTR), the sensor is thus read continuously, with a full frame reading time of 1.3\,s, producing a ramp of accumulating signal. A single, cold optical element (the L12 cold lens) is placed in front of the sensor and is the final element of the optical path between DDRAGO and CAGIRE.

To allow the sensor to operate at 100\,K, it is placed along with its pre-amplifier within a cryostat especially developed. A secondary vacuum ($1.5 \times 10^{-7}$\,mbar) is first achieved using a vacuum pump. Then, cryocooling lowers the temperature to 100\,K and maintains the vacuum via cryogenic pumping. A cold finger (80\,K) distributes the cold inside the cryostat using multiple thermal braids, cooling both the sensor and the baffle alongside the optical path to lower internal thermal emission. A cold shutter can also be actuated to cover the entry pupil inside the cryostat for characterization purposes. To maximise the life expectancy of the ALFA sensor, the cryostat will remain cooled at 100\,K, only to be temporarily warmed up for annealing the sensor if necessary (once per year or less).

\begin{figure}
    \centering
    \includegraphics[width=0.49\linewidth]{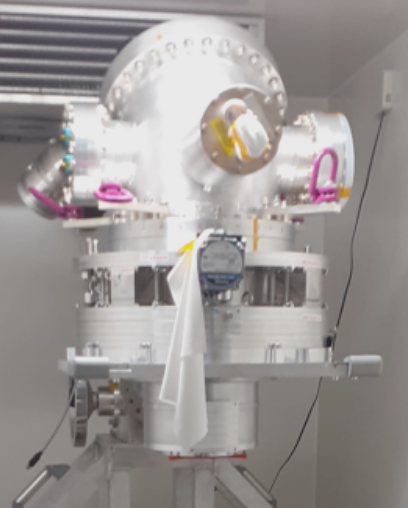}
    \includegraphics[width=0.49\linewidth]{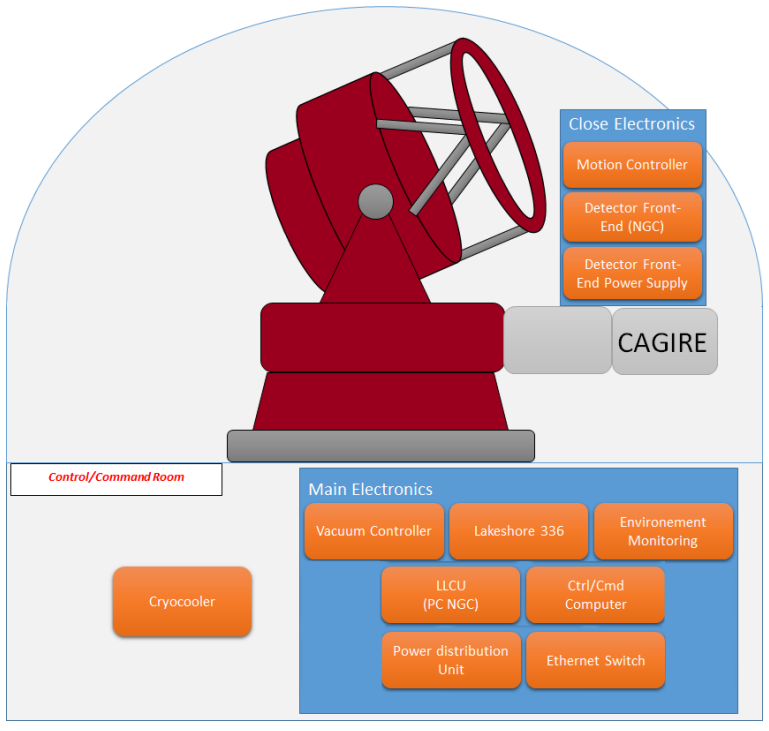}
    \caption{The CAGIRE camera fully assembled at IRAP (left) and setup within the observatory (right). Proximity electronics are needed to ensure minimal electrical noise, adding to the mechanical constraints. Main electronics and cryogenics are located in the lower control room.}
    \label{fig:CAGIRE}
\end{figure}

The astronomical $J$ and $H$ filters are located outside the cryostat on a motorized filter rack in front of the entry pupil. The body of the cryostat is equipped with a manual alignment stage, which allows movement in translation (2 axes) and rotation (3 axes). Optical alignment is performed once at installation and locked in place to achieve common relative focus and field of view with the DDRAGO camera, a moving lens inside the WOB permits the precise focal adjustment of the NIR sensor.

\subsubsection{Characterization and Performances}

The camera has been assembled and the ALFA sensor successfully integrated in summer 2025 at IRAP, France. Coupled to the New Generation Controller (NGC) provided by ESO, it delivers infrared images of excellent quality when cooled to the operating temperature. Notably, the noise level of dark images taken with the camera is compatible with the noise level measured by CPPM, France on the sensor in a controlled environment, which confirms the quality of the thermal and electrical design as well as the successful integration.

As of autumn 2025, CAGIRE is undergoing a battery of tests to validate its characteristics --- in the dark, under illumination, at various external temperatures, tilted in different directions --- and confirm a stable delivery of high-quality images in realistic observing conditions mimicking the ones at the observatory. It is planned for installation in Spring 2026, where control and acquisitions can be tested before the Warm Optical Bench (WOB) is manufactured for a first light at the end of 2026.

\subsubsection{Pre-processing software}

The raw data output from a single exposure of CAGIRE consists of a set of frames of accumulated signal, taken every 1.3\,s. However, the product that must be delivered is a single image containing the measured flux and associated uncertainty for each pixel. Hence, after each acquisition N, a pre-processing software will partially reduce the raw data output, and provide a single image before the end of the next acquisition, to ensure a smooth operation. 

The software computes for each sensitive pixel a differential ramp to  estimate the average flux during acquisition. The software handles non-linearity, cosmic ray impacts and saturated pixels by fitting only parts of the ramps that are not affected, increasing the effective dynamic range of the sensor at the cost of slightly increased noise in concerned pixels.

\begin{figure}
    \centering
    \includegraphics[width=1.0\linewidth]{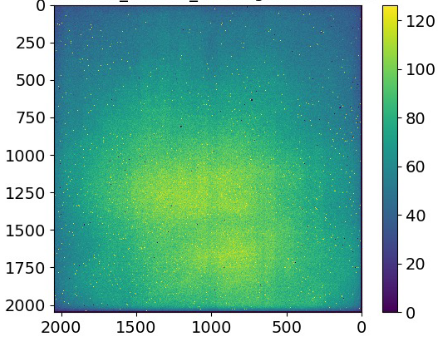}

    \caption{Pre-processed CAGIRE image of a uniform blackbody located in front of the entrance window. sensor cosmetics are excellent, and flux response is as expected.}
    \label{fig:CAGIRE_image}
\end{figure}

\subsubsection{Observational strategy and on-sky performances}

When reacting to a live alert, the default filter will be set to $H$-band. This is in order to maximise the chances of catching the afterglow of a highly-redshifted or absorbed event. If a counterpart is identified in $H$-band, and/or an optical counterpart is detected with DDRAGO, the observation blocks can be set to alternating $J$- and $H$-band exposures. Additionally, with both $J$- and $H$-band measurements, the photometric redshift capabilities of COLIBRÍ are extended up to $z\approx8$ with a 10\%\, uncertainty \citep{2018SPIE10705E..1RC}.

By combining an end-to-end image simulator of the camera with the results from the characterisation of the ALFA sensor by CEA and CPPM in France, we have evaluated the performances of CAGIRE in realistic scenario of far away GRB events. Of the nine high-redshift events with $z>6$ that have occurred in the past 20 years, four would have been detected. We then tested the limits of CAGIRE by artificially placing those four events at greater redshifts, which results in the detection of their afterglow up to $z \approx 13$ \citep{2024A&A...691A.324F}, distance at which the Lyman break is shifted outside the H-band.

\subsection{Telescope Control System}
\label{subsec:obs-control}

The Telescope Control System (unimaginatively called TCS) is a development of the system used by RATIR \citep{butler-2012,watson-2012}, COATLI \citep{watson-2016a}, and DDOTI \citep{watson-2016b}. The system actually manages all aspects of observatory, telescope, and instrument observations prior to data being written to disk for analysis by the pipeline, including determining safe conditions for opening, scheduling, alert management, normal guest-observer block management, focusing, pointing correction, executing alerts and blocks, monitoring, and notification.

The system was originally written in Tcl, as that was the language favored by the Observatorio Astronómico Nacional when the RATIR project began in 2008. However, for its deployment on COLIBRÍ, all configuration files and inter-process communication were moved to language-neutral formats (JSONH and JSON-RPC). This has allowed new developments to be written in Python, which is much better supported today.

Adapting TCS to COLIBRÍ required writing a new code to interface with the ASTELCO OpenTSI telescope controller (1000 lines) and with the Spectral Instruments Windows-based CCD controllers (1000 lines). Other developments were web-based interfaces for both the alert queue and guest-observer blocks. The alert queue interface allows on-shift scientists to easily manage alerts, for example, updating coordinates if an optical counter-part is reported in a GCN Circular or changing the filters according to the magnitude of a new transient.

\subsection{Data Processing}

The COLIBRI image analysis pipeline was developed based on the DDOTI, COATLI and RATIR pipelines~(e.g. \citet{2023MNRAS.525.3262B,2018ApJ...857...81G}). It is an end-to-end processing framework designed to transform raw telescope images into calibrated light curves and to identify transient sources within the observations. The system is orchestrated by a Python controller and integrates modules written in Python, C, and Bash. 

\begin{figure*}
    \centering
   \includegraphics[height=0.52\textwidth]{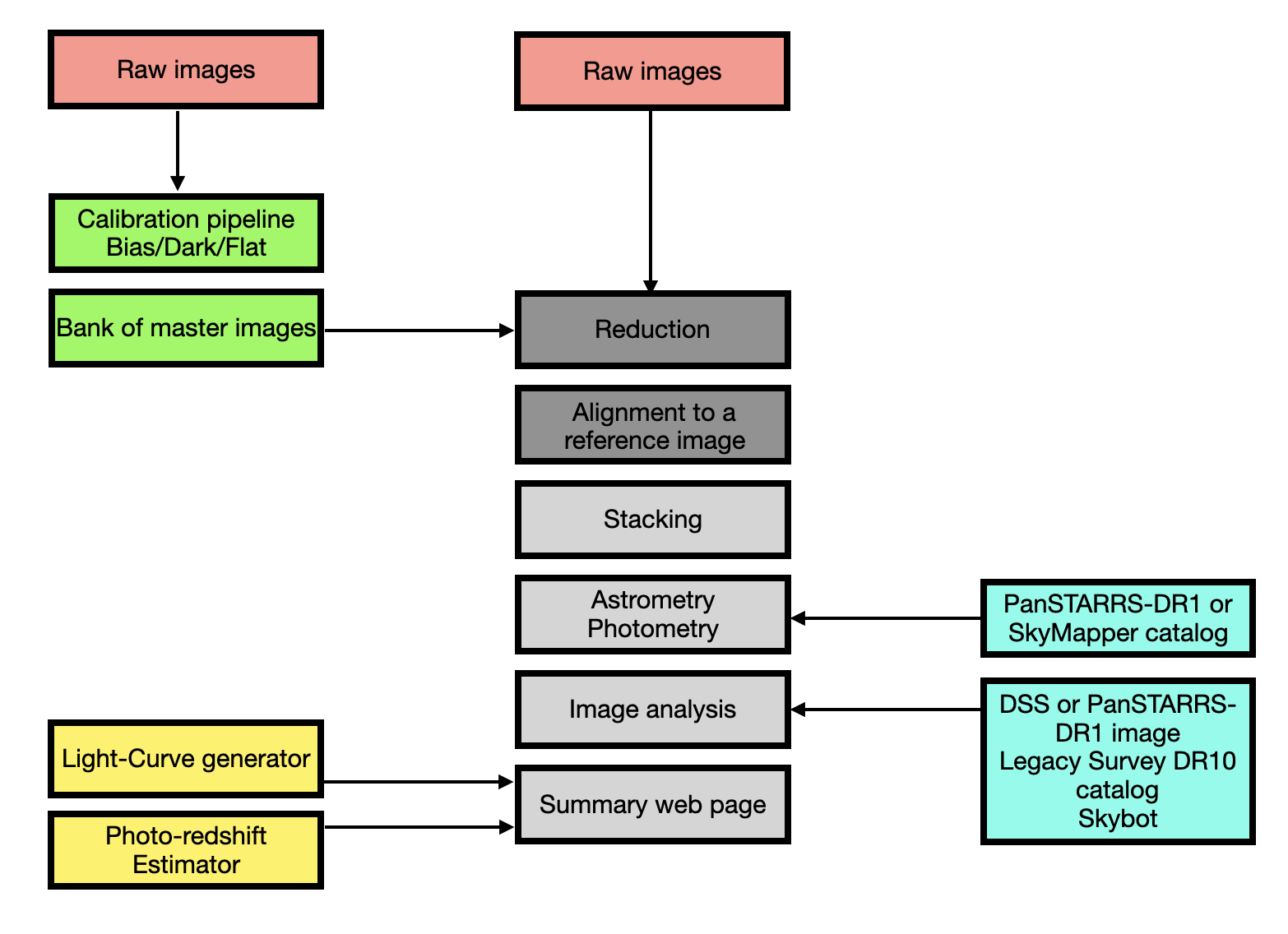}
    \caption{Scheme of the data-processing pipelines used for the COLIBRI image analysis. In green, it is illustrated the calibration pipeline run every day while the steps of the analysis pipeline are shown as grey boxes. The two first steps are processed on fly in the online mode, the others are run after a given number of reduced images. The blue boxes refer to external inputs needed for the astrometry, photometry and analysis steps. The two yellow boxes correspond to two auxiliary tools that draw the lightcurve and estimate the photo-redshit.}
    \label{figure:pipeline}
\end{figure*}

Its modular design allows users to adapt the analysis workflow to their specific observing conditions. 
The pipeline operates through a series of stages (see Figure~\ref{figure:pipeline}): \textit{reduction}, \textit{alignment}, \textit{stacking}, \textit{photometric/astrometric calibration}, and \textit{analysis}. \\

\subsubsection{Reduction}

Image reduction begins with the creation of master calibration frames—dark, bias, and flat-field images—generated from corresponding input frames after a quality assessment based on their mean, median, and standard deviation (calibration pipeline). 

The reduction process then applies standard corrections for dark, bias, and flat-field effects, removes defective or hot pixels, corrects image rotation, and produces a background weight map. \\

\subsubsection{Alignment}

Alignment is performed by extracting sources from each image using SExtractor \citep{Bertin1996}. 
A reference image is selected as the median-quality frame, determined from the number of detected stars. Images with too few stars (noisy or poor quality) or too many stars (saturated or contaminated) are automatically discarded. 

Source positions are cross-matched with the reference image to determine precise shifts. 
The alignment is refined iteratively by rejecting outliers and applying thresholds on stellar parameters such as the Full Width at Half Maximum (FWHM). 

This systematic procedure ensures robust astrometric alignment and high-quality inputs for the subsequent steps. 

\subsubsection{Stacking}

Stacking is performed using the SWarp tool \citep{Bertin2002}, which resamples and combines images onto a common reference grid. 

An initial median stack is generated to minimize the influence of noise and artifacts. Each individual image is then compared to this median stack to identify residual artifacts and produce dedicated masks for the final combination. 

The final stacked image is produced using a weighted average with iterative intermediate stacks, the number of which is user-configurable. Sky subtraction is performed using a Lagrange polynomial fit.

\subsubsection{Photometry and Astrometry calibration}

Photometric calibration begins with astrometric calibration, performed using the astrometry.net suite~\citep{Lang2010}. Aperture photometry is carried out with SExtractor, using an aperture diameter of $2 \times \mathrm{FWHM}$, where FWHM is estimated directly from the source extraction. 

Photometric zero points are derived using Pan-STARRS DR2 ~\citep{2016arXiv161205560C} for the Northern hemisphere (declination $> -28^\circ$) and SkyMapper DR4 ~\citep{2024PASA...41...61O} for the Southern hemisphere. The zero point can be modeled either as a single constant value or as a polynomial function of the pixel coordinates. 

\subsubsection{Analysis}

The analysis module processes the stacked field images to identify transient candidates and generate their light curves. Sources present both in the reference catalog (Pan-STARRS or SkyMapper) and in the stacked image are flagged as matched, while unmatched detections are classified as potential transient candidates. 

Matched sources showing significant photometric deviations relative to the catalog magnitude are also flagged as variable or interesting. For all transient candidates, light curves are produced using first-half and second-half image stacks to estimate temporal slopes.

\subsubsection{Online processing}

To meet the specific requirements of rapid GRB afterglow identification in COLIBRI observations, a dedicated online mode has been developed. This mode is optimized for low-latency processing, producing scientific products (stacked images, calibrated photometry, transient candidates) within minutes of the first exposure. 

The pipeline runs automatically at the telescope site, processing each image as it is acquired. 
Once a predefined number of images have been obtained, the stacking, photometry, and analysis steps are triggered. The results are displayed on dedicated web pages and made available in real time to the observers on duty.

\subsubsection{Cross-check and multi-camera analysis}

The stdweb tool~\citep{2024arXiv241116470K} is used to cross-check transient detections and verify photometric consistency. A multi-camera analysis module has also been developed to construct multi-band light curves for each candidate and estimate its photometric redshift via Spectral Energy Distribution (SED) fitting~\citep{2018SPIE10705E..1RC}. 

The redshift is derived using a Markov Chain Monte Carlo minimization of the model SED at a given epoch, with extrapolation applied when the multi-band data are not strictly simultaneous. 
The intrinsic GRB afterglow emission is modeled as a single power law from optical to near-infrared wavelengths, yielding four fitted parameters: the photometric redshift $z$, the dust extinction coefficient $A_V$ in the $V$~band, the spectral index $\beta$, and a normalization factor norm used to scale the flux. This procedure has been validated using simulated GRB afterglows and archival observations.

\section{Performance in the follow-up of SVOM alerts}

COLIBRÍ’s scientific observations effectively began in early January 2025. Over the span of 10 months, approximately 140 GCNs have been issued, with the following distribution:

\begin{itemize}
\item 36\% for SVOM,
\item 36\% for Swift,
\item 22\% for Einstein Probe,
\item the remaining fraction for IceCube, CHIME, etc.
\end{itemize}

Approximately 71\% of the SVOM alerts have been followed by COLIBRI during this period, which is perfectly in line with the initial expectations (about 65\%). If we exclude alerts that are in fact not observable (i.e. located far in the south), it gives an efficiency in the follow-up of SVOM alerts of about 77\%. This efficiency is excellent and, in fact, the loss of 23\% is mainly due to unfavorable weather conditions\footnote{The median seeing measured by our seeing monitor,  0\farcs72 , combined with the also measured very high fraction of usable nights, ~80\%, clearly demonstrates that the Observatorio Astronómico Nacional in the Sierra de San Pedro Mártir, Baja California, Mexico, ranks among the best in the world for astronomical observations.}.

The detection efficiency was estimated after verifying the validity of the SVOM/ECLAIRs triggers \citep{Godet+etal+2026a}. Between 65\% and 71\% of the GRBs were successfully detected by COLIBRI, while the remaining 35\% to 29\% resulted in upper limits only. These figures are fully consistent with the initial expectations, and they confirm that the telescope and its automation pipeline deliver the level of performance required for routine follow-up of high-energy transients.

The response time is also excellent, with an average delay of about 60 seconds between the SVOM onboard trigger and the start of the COLIBRÍ observation, in cases where the alert is observable in real time. This delay covers the entire chain, from the trigger on board to transmission through the VHF network, interruption of the ongoing observation, and initiation of the observation sequence associated with the alert.

\section{Conclusion}
COLIBRÍ aims to play a key role in the follow-up of GRBs detected by SVOM, as well as in the study of the most transient astrophysical phenomena, such as FRBs, FBOTs, optical counterparts of gravitational waves and high-energy neutrinos, supernovae, and more. The system offers an excellent compromise between rapid response and sensitivity, with simultaneous observations in multiple bands from the visible to the near-infrared, while remaining technically manageable thanks to its reasonable telescope size. The first results obtained in 2025 place COLIBRÍ among the most efficient facilities of its kind,  validating the design choices made during its development.

Although the project is still young, it is expected to reach its full potential by the end of 2026 with the commissioning of the near-infrared camera, CAGIRE. In parallel, the team is already working actively on a second instrument for the additional Nasmyth focus: a polarimeter named \textit{TEQUILA}, which would provide unique scientific capabilities.

Once equipped with both instruments, COLIBRÍ will be fully operational and will provide very high-quality data to the SVOM mission, as well as to the French and Mexican scientific communities.

\hrule

\begin{acknowledgements}
COLIBRÍ is funded by the Universidad Nacional Autónoma de México (CIC and DGAPA/PAPIIT IN109418 and IN109224), and CONAHCyT (1046632 and 277901). It received financial support from the French government under the France 2030 investment plan, as part of the Initiative d’Excellence d’Aix-Marseille Université-A*MIDEX (ANR-11-LABX-0060 – OCEVU and AMX-19-IET-008 – IPhU), from LabEx FOCUS (ANR-11-LABX-0013), from the CSAA-INSU-CNRS support program, and from the International Research Program ERIDANUS from CNRS.
The ALFA detector is property of ESA, we thank ESA and the LabEx FOCUS for making it available for CAGIRE. The view expressed herein can in no way be taken to reflect the official opinion of the European Space Agency.
JLA and HV thank F.~Beigbeder, P.~Couderc and P.~Amber for their contribution on the CAGIRE prototype.
The COLIBRí and CAGIRE teams thank D. Barret for his positive support in the early phases of the project.
\end{acknowledgements}

\label{lastpage}

\clearpage
\bibliography{ms2026-0016}{}
\bibliographystyle{raa} 

\end{document}